\documentclass[journal]{IEEEtran}
\usepackage{cite}
\ifCLASSINFOpdf
   \usepackage[pdftex]{graphicx}
\else
   \usepackage[dvips]{graphicx}
\fi

\usepackage{multirow}
\usepackage{amsmath}
\usepackage{steinmetz}
\usepackage{color}
\usepackage{amssymb}

\newcommand{\romdig}[1]{\MakeUppercase{\romannumeral #1}}

\ifCLASSOPTIONcompsoc
 \usepackage[caption=false,font=normalsize,labelfont=sf,textfont=sf]{subfig}
\else
 \usepackage[caption=false,font=footnotesize]{subfig}
\fi

\begin{document}

\title{Multifrequency System Model for Multiport Time-Modulated Scatterers}

\author{Aleksandr~D.~Kuznetsov,~\IEEEmembership{Student~Member,~IEEE},
        Jari~Holopainen,
        and~Ville~Viikari,~\IEEEmembership{Senior~Member,~IEEE}
\thanks{Aleksandr~D.~Kuznetsov,~Jari~Holopainen,~and~Ville~Viikari are 
        with the Department of Electronics and Nanoengineering, Aalto University, Espoo, Finland (e-mail: aleksandr.kuznetsov@aalto.fi).}
\thanks{This work has been submitted to the IEEE for possible publication. Copyright may be transferred without notice, after which this version may no longer be accessible. \emph{(Corresponding author: Aleksandr~D.~Kuznetsov)}}}


\maketitle

\begin{abstract}

Utilizing scatterers in communication engineering, such as reconfigurable intelligent surfaces (RISs) and backscatter systems, requires physically consistent models for accurate performance prediction. A multiport model, which also accounts for structural scattering, has been developed for non-periodic scatterers. 
However, many emerging systems operate at multiple frequencies or generate intermodulation harmonics, particularly when incorporating space-time modulation (STM) or dynamic load control. These functionalities demand advanced modeling approaches capable of capturing scattering behavior across several frequencies and directions simultaneously.
This article extends a multiport S-parameters-based model for predicting the scattering properties of multifrequency operating structures. The model extends the applicability of convenient S-matrix models to time-modulated multiport structures. 
Unlike known approaches, this model incorporates structural scattering, mutual coupling, the possibility of non-digital modulation, and non-periodic configurations, enabling precise analysis and optimization for a broad range of communication and sensing systems. 
Validation against experimental results for a space-time modulated scattering structure demonstrates the accuracy and practical applicability of the proposed model.

\end{abstract}


\IEEEpeerreviewmaketitle

\section{Introduction}

\IEEEPARstart{F}{or previous} several years, an interest in the functionality of scattering systems has been invoked, especially in communication engineering. 
The utilization of adjustable scatterers forms the foundation of smart electromagnetic environments (SEME) (also referred to as smart radio environments in some sources). Their primary purpose is to enhance the communication channel by guiding the transmitting signal to the receiver \cite{smart_env}.
Scatterers can be applied as passive static reflective skins or reconfigurable intelligent surfaces (RISs) in cases where they do not cause time modulation of the scattered signals but enable signal redirection to a desired direction \cite{smart_env,table_SEME}. 
Oliveri et al. \cite{new_paradigm_S_istead_D} demonstrated that optimizing these scatterers and associated systems based on the communication quality of all components collectively (capacity-driven methods) provides a more accurate practical approach than relying on optimization of the free-space performance of individual elements (directivity-based methods). 
Since scattering parameters (S-parameters) can be connected to channel efficiency characteristics, several models utilizing S-parameters or related impedances were proposed to describe behavior of scatterers either as part of communication channel \cite{ris_communication,Renzo_S_param_opt_fw_2024,scat_dipoles_theory,Z_to_comm_with_str_2024,ZS_2024,Z_param_optim,our_S_param,comm_eng_low_scat,Sravan_model_part_2_channel,beyond_diagonal_RIS,beyond_diagonal_RIS_channels} or as external to it \cite{our_S_param,modal_S_ant_syst,S_ant_meas,Sravan_model_part_1_structure}, depending on the load impedance values. Additionally, the S-parameters-based approach is applicable to traditional backscatter communication systems and energy harvesting analysis \cite{backscatter_types,general_BackCom}.

\begin{figure}[!t]
\centering
\includegraphics[width=\columnwidth]{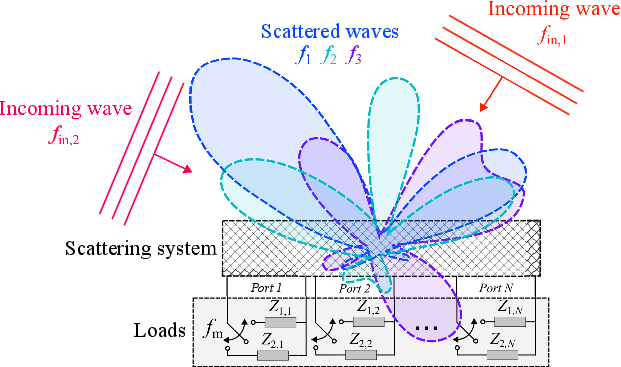}
\caption{Schematic view of the multiport system model for a multifrequency operating scattering structure description.}
\label{General_scat_scheme}
\end{figure}

Alternatively, scatterers can be applied to process signals (Fig.~\ref{General_scat_scheme}), for example, by acting as a smart repeater -- which is a non-regenerative relay able to amplify and forward signals -- or RIS in SEME \cite{smart_env,table_SEME,lit_review_ris}, or in ambient backscatter systems \cite{backscatter_types}. One highly functional solution is the space-time-modulated (STM) structure, with parameters varying dynamically in space and time \cite{STM_Base_theory,smart_env,STMTM_part_1,STMTM_part_2,lit_review_ris,S_TM_array,S_TM_ph_array,S_TM_communication}. 
A prominent example is a space-time-coding (STC) digital metasurface (DM), composed of unit elements whose time-domain control enables simultaneous beam steering and harmonic power shaping \cite{Space_time_MTS}. STC DMs and arrays have been used for wavefront shaping, harmonic generation, reciprocity breaking, and polarization manipulation \cite{STC_base,STC_applications,STC_polarization,MTM_thory_and_prac_zero_2,STC_nonreciprocity}, and applied in sensing \cite{STC_sensing}, pattern modification, signal modulation, multiplexing, MIMO and RIS systems \cite{STC_applications,STC_modulator,Alamouti_STM_RIS,ML_STCDM_15dBdiff}.

Based on \cite{Space_time_MTS,STC_base}, most STC DM modeling approaches have constraints associated with structural periodicity. These include the inability to describe aperiodical structures, the neglect of mutual coupling between unit elements, and the assumption of isotropic behavior of the unit elements \cite{Space_time_MTS,STC_applications,MTM_thory_and_prac_zero_2,simovski_tretyakov_2020}. Furthermore, many models omit structural scattering effects, restrict the load modulation to digital states only, and ignore variations in reflection amplitude between coding states, which limits their accuracy in practical applications \cite{struct_scat_importance,STC_applications}.

While STM scatterers are typically described using periodic, sub-wavelength unit cells, to the best of our knowledge, they have not been analyzed within a multiport scattering parameter model (Fig.~\ref{General_scat_scheme}), which enables channel characterization and optimization, taking into account coupling and structural scattering \cite{Renzo_S_param_opt_fw_2024,new_paradigm_S_istead_D,our_S_param,Z_param_optim,comm_eng_low_scat}. This model becomes particularly critical for systems or angular configurations where the metasurface approach cannot be used for an accurate description.

In the conventional definition, S-parameters are not applicable for describing the behavior of systems containing multiple tones or producing frequency conversions~\cite{Pozar_2011,X_parameters_book_2013}. To address this limitation, the polyharmonic distortion (PHD) model \cite{PHD_model,nonlinear_circuits_book_2018,X_parameters_book_2013} and the multitone multiharmonic scattering parameters (M$^2$S) \cite{m2s_expand} were developed. Initially designed for systems with active components, these models can be used for multitone, multiport systems, enabling the characterization of harmonic interactions, including intermodulation effects \cite{X_parameters_book_2013,nonlinear_circuits_book_2018,m2s_expand}. Applying an analogical approach to scattering systems should provide a theoretical framework for describing their broader functionality, including STM scatterers.

This article presents a model for the numerical characterization of multistatic scattering in arbitrarily loaded scattering structures operating at multiple frequencies. To achieve this, the scatterer is described as a multiport system by extending the conventional S-parameters approach into the multifrequency model, enabling simultaneous evaluation of scattering behavior across multiple frequencies while retaining the practical benefits of the S-parameters framework. Unlike existing models, this approach explicitly accounts for mutual coupling and variations in the unit element patterns, without relying on periodicity assumptions or homogenization, and without neglecting the resistive part of the loading circuits.
The key novelty of the model lies in opening a possibility to predict multifrequency scattering behavior while maintaining physical consistency and model simplicity. Moreover, the model can use analytical, simulated, or measured data as its input. This capability gives an opportunity for the development and optimization of scatterers for diverse practical applications.
To validate the effectiveness of the proposed model, a space-time-modulated scattering structure was fabricated, measured, and analyzed under varying operational regimes, demonstrating both the accuracy and practical applicability of the method.

This article is organized as follows. Section~\ref{sec:theory} expands the multiport S-parameters model to predict scattering with multiple input tones and over generated harmonic frequencies. Section~\ref{sec:experiment} describes the structure and measurement setup of experimental confirmation of the proposed model. Section~\ref{sec:results} compares the computed and measured results and analyzes the performance of the proposed model for the STM case. Finally, Section~\ref{sec:conclusion} concludes the paper.

\section{Multifrequency Multiport System Model}
\label{sec:theory}

In this section, we present a multifrequency multiport scattering model to predict the scattering of a structure terminated with arbitrary modulated loads. Analogous to the model presented in \cite{our_S_param}, the full scattering matrix is expressed as a combination of submatrices, separated based on their physical sense. Here, we extend the S-parameters approach to enable multifrequency characterization of scatterers terminated with modulated loads, detailing the formation of submatrices and establishing their relation to BCS values.

\subsection{Idea of the Model}

\begin{figure}[!t]
\centering
\includegraphics[width=\columnwidth]{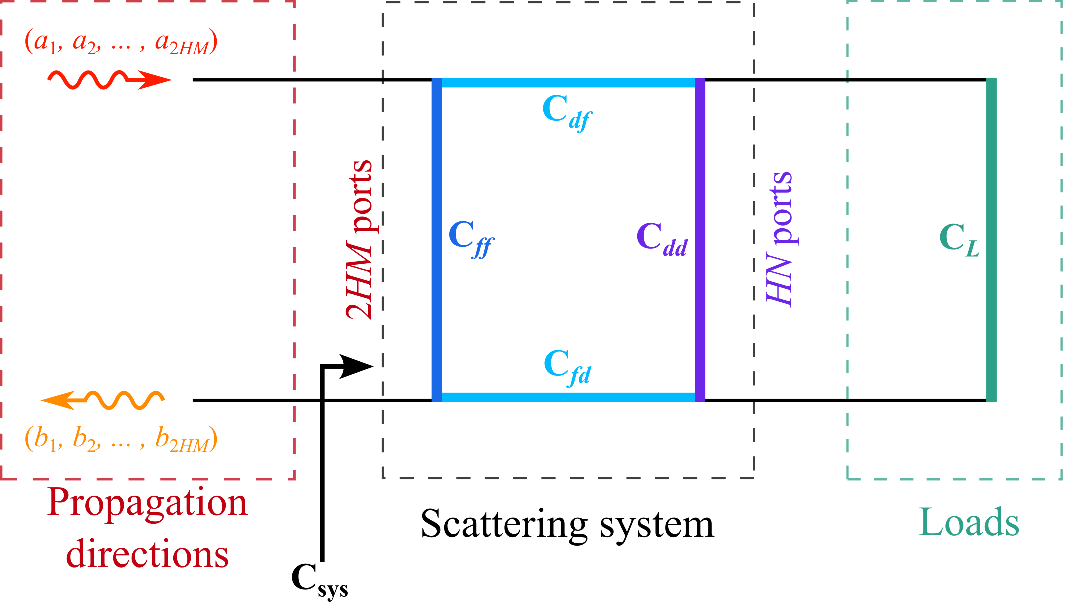}
\caption{The proposed multifrequency model for scatterer description, an extension of the S-parameters approach in \cite{our_S_param}.}
\label{C_scheme}
\end{figure}

The proposed S-parameters model for scattering structure description was developed in \cite{Renzo_S_param_opt_fw_2024,S_ant_meas} and extended in \cite{our_S_param} to enable simultaneous characterization of scattering in multiple directions. In this approach, S-parameters submatrices connect $M$ radiation ports -- corresponding to the number of propagation directions under consideration -- and $N$ discrete load ports.
Building upon this framework, we further develop the model to account for multifrequency operation.
However, scattering behavior across several frequencies cannot be described using standard S-parameters \cite{X_parameters_book_2013}. To address this, each radiation port in the developed model is separated into $H$ subports, where $H$ denotes the number of frequency harmonics under consideration. Similarly, each load port is separated into $H$ subports to account for harmonic coupling at the loads. The parameter $H$ is selected under the assumption of a discrete spectrum of the scattered signal.
While ideally $H\to\infty$ would describe distortions across all generated harmonics, in practice, a sufficiently large value of $H$ ensures accurate modeling. 
Additionally, the model is extended to account for polarization by connecting polarization components, resulting in $2HM$ radiation subports. Each radiation subport thus refers to a particular polarization component of a signal harmonic at one of $H$ frequencies in one of $M$ directions.
A schematic representation of the developed model is shown in Fig.~\ref{C_scheme}.

Redefined matrices supporting the coupling between the frequencies we denote as $\relax[\mathbf{C}]$. These matrices are based on the analogical S-parameter matrices $\relax[\mathbf{S}]$ described in \cite{our_S_param}. Thus, the connection between newly defined ports, encapsulating both polarizations, is possible:
\begin{equation}
\begin{bmatrix}
    [\mathbf{b}^{\varphi}]_1\\
    [\mathbf{b}^{\theta}]_1\\
    [\mathbf{b}^{\varphi}]_2\\
    \vdots \\
    [\mathbf{b}^{\theta}]_H
\end{bmatrix}
=
\relax[\mathbf{C_\text{sys}}]
\begin{bmatrix}
    [\mathbf{a}^{\varphi}]_1\\
    [\mathbf{a}^{\theta}]_1\\
    [\mathbf{a}^{\varphi}]_2\\
    \vdots \\
    [\mathbf{a}^{\theta}]_H
\end{bmatrix},
\label{eq:abC}
\end{equation}
where $[\mathbf{a}^{\chi}]_h$ and $[\mathbf{b}^{\chi}]_h$ are $M \times 1$ vectors of input and output power waves \cite{Pozar_2011} defined for $\chi$ polarization and $h$ harmonic, correspondingly. Analogous to \cite{our_S_param},
\begin{equation}
    \mathbf{C_\text{sys}}=\mathbf{C_\emph{ff}}
    +\mathbf{C_\emph{fd}}\mathbf{C_\emph{L}}(\mathbf{I}-\mathbf{C_\emph{dd}\mathbf{C_\emph{L}}})^{-1}\mathbf{C_\emph{df}},
\label{eq:C_sys}
\end{equation}
where matrix $\relax[\mathbf{C_\emph{ff}}]$ connects radiation ports, $\relax[\mathbf{C_\emph{fd}}]$ and $\relax[\mathbf{C_\emph{df}}]$ connect radiation and load ports, $\relax[\mathbf{C_\emph{dd}}]$ represents interactions among discrete ports via the scattering structure, $\relax[\mathbf{C_\emph{L}}]$ characterizes the load network of the structure, and $\mathbf{I}$ is the identity matrix.
It is important to note that this representation is valid only under the assumptions outlined in the following subsections.

\subsection{Formation of Matrices Based on Structures of Antennas}

The following text describes the process of forming all $\relax[\mathbf{C}]$ matrices except the load one. We can consider each load port terminated by a unit antenna scattering element to connect the radiation and load ports. Because antennas are linear systems, frequency harmonics differing from those injected into the system can only be generated by the loads. Thus, all $\relax[\mathbf{C}]$ matrices except the load one are derived from the corresponding $\relax[\mathbf{S}]$ matrices, as described in \cite{our_S_param}, establishing connections only between harmonics having identical indices. Further, the superscript index denotes the polarizations or transitions between polarizations, as defined by the respective parameter. Additionally, all S-parameter matrices are normalized to the same reference impedance $Z_\text{ref}$.

The $2HM \times 2HM$ matrix $\relax[\mathbf{C_\emph{ff}}]$, describing the structural scattering of the system, may be formed as

\begin{equation}
\relax[\mathbf{C_\emph{ff}}] =
\begin{bmatrix}
    [\mathbf{S}_\emph{ff}]_1 & [\mathbf{\emph O}]_{2M,2M} & \hdots & [\mathbf{\emph O}]_{2M,2M}\\
    [\mathbf{\emph O}]_{2M,2M} & [\mathbf{S}_\emph{ff}]_2 & \hdots & [\mathbf{\emph O}]_{2M,2M}\\
    \vdots & \vdots & \ddots & \vdots\\
    [\mathbf{\emph O}]_{2M,2M} & [\mathbf{\emph O}]_{2M,2M} & \hdots & [\mathbf{S}_\emph{ff}]_H
\end{bmatrix},
\label{eq:C_ff}
\end{equation}
where $[\mathbf{\emph O}]_{2M,2M}$ is a $2M \times 2M$ zero matrix, and the unified $\relax[\mathbf{S_\emph{ff}}]_h$ matrix formed for each harmonic $h$ is
\begin{equation}
    \relax[\mathbf{S_\emph{ff}}]_h =
\begin{bmatrix}
    [\mathbf{S}_\emph{ff}^{\varphi \rightarrow \varphi}]_h & [\mathbf{S}_\emph{ff}^{\theta \rightarrow \varphi}]_h\\
    [\mathbf{S}_\emph{ff}^{\varphi \rightarrow \theta}]_h &  [\mathbf{S}_\emph{ff}^{\theta \rightarrow \theta}]_h
\end{bmatrix}.
\label{eq:S_ff}
\end{equation}
Each of the constituent matrices is the $M \times M$ matrix that connects radiation subports at harmonic $h$ and includes scattering through linear external channels, channel parts other than the scatterer (e.g., radiating antenna), and structural scattering of the scatterer itself.

Acting identically for the $2HM \times HN$ matrix $\relax[\mathbf{C_\emph{fd}}]$ that connects the radiation and load ports,
\begin{equation}
\relax[\mathbf{C_\emph{fd}}] =
\begin{bmatrix}
    [\mathbf{S}_\emph{fd}]_1 & [\mathbf{\emph O}]_{2M,N} & \hdots & [\mathbf{\emph O}]_{2M,N}\\
    [\mathbf{\emph O}]_{2M,N} & [\mathbf{S}_\emph{fd}]_2 & \hdots & [\mathbf{\emph O}]_{2M,N}\\
    \vdots & \vdots & \ddots & \vdots\\
    [\mathbf{\emph O}]_{2M,N} & [\mathbf{\emph O}]_{2M,N} & \hdots & [\mathbf{S}_\emph{fd}]_H
\end{bmatrix},
\label{eq:C_fd}
\end{equation}
where $[\mathbf{\emph O}]_{2M,N}$ is a $2M \times N$ zero matrix and the unified $\relax[\mathbf{S_\emph{fd}}]_h$ matrix formed for each harmonic $h$ is
\begin{equation}
    \relax[\mathbf{S_\emph{fd}}]_h =
\begin{bmatrix}
    [\mathbf{S}_\emph{fd}^{\varphi}]_h\\
    [\mathbf{S}_\emph{fd}^{\theta}]_h
\end{bmatrix},
\label{eq:S_fd}
\end{equation}
consisting of two $M \times N$ matrices connecting radiation and load subports. For example, this connection can be based on radiation patterns of unit elements, which include possible mismatch of the antenna (having impedance $Z_a(f)$), transmission lines (having impedance $Z_t(f)$), and the load port \cite{our_S_param}.

Following the same physical principles,
\begin{equation}
\relax[\mathbf{C_\emph{df}}] =
\begin{bmatrix}
    [\mathbf{S}_\emph{df}]_1 & [\mathbf{\emph O}]_{N,2M} & \hdots & [\mathbf{\emph O}]_{N,2M}\\
    [\mathbf{\emph O}]_{N,2M} & [\mathbf{S}_\emph{df}]_2 & \hdots & [\mathbf{\emph O}]_{N,2M}\\
    \vdots & \vdots & \ddots & \vdots\\
    [\mathbf{\emph O}]_{N,2M} & [\mathbf{\emph O}]_{N,2M} & \hdots & [\mathbf{S}_\emph{df}]_H
\end{bmatrix},
\label{eq:C_df}
\end{equation}
where $[\mathbf{\emph O}]_{N,2M}$ is a $N \times 2M$ zero matrix and the unified $\relax[\mathbf{S_\emph{df}}]_h$ matrix formed for each harmonic $h$ is
\begin{equation}
    \relax[\mathbf{S_\emph{df}}]_h =
\begin{bmatrix}
    [\mathbf{S}_\emph{df}^{\varphi}]_h
    [\mathbf{S}_\emph{df}^{\theta}]_h
\end{bmatrix}.
\label{eq:S_df}
\end{equation}

The $HN \times HN$ matrix $\relax[\mathbf{C_\emph{dd}}]$ describes the coupling between scattering antennas at the frequencies under consideration and is polarization-independent
\begin{equation}
\relax[\mathbf{C_\emph{dd}}] =
\begin{bmatrix}
    [\mathbf{S}_\emph{dd}]_1 & [\mathbf{\emph O}]_{N,N} & \hdots & [\mathbf{\emph O}]_{N,N}\\
    [\mathbf{\emph O}]_{N,N} & [\mathbf{S}_\emph{dd}]_2 & \hdots & [\mathbf{\emph O}]_{N,N}\\
    \vdots & \vdots & \ddots & \vdots\\
    [\mathbf{\emph O}]_{N,N} & [\mathbf{\emph O}]_{N,N} & \hdots & [\mathbf{S}_\emph{dd}]_H
\end{bmatrix}.
\label{eq:C_dd}
\end{equation}
Here, $[\mathbf{\emph O}]_{N,N}$ is a $N \times N$ zero matrix and $\relax[\mathbf{S_\emph{dd}}]_h$ is a $N \times N$ matrix connecting discrete ports formed for each harmonic $h$.

It is important to note that all the formation processes for matrices $\relax[\mathbf{C_\emph{ff}}]$, $\relax[\mathbf{C_\emph{fd}}]$, $\relax[\mathbf{C_\emph{df}}]$, $\relax[\mathbf{C_\emph{dd}}]$ can be based on theoretical, simulation, or measurement data as a source of information \cite{our_S_param}. 
Matrices are based on the assumption of linear behavior of the scatterer and alternative channels (in case of $\relax[\mathbf{C_\emph{ff}}]$) at each considered frequency, utilizing the superposition principle for the signals, which is useful for the description of periodically-modulated systems.

Unlike the described matrices, the $HN \times HN$ load matrix $\relax[\mathbf{C_\emph{L}}]$ should describe the behavior of time-modulated loads which can cause frequency shifts to the signals. Therefore, the corresponding load matrix is constructed differently.

\subsection{Example of Modulation by Switched Loads}

\begin{figure}[!t]
\centering
\includegraphics[width=\columnwidth]{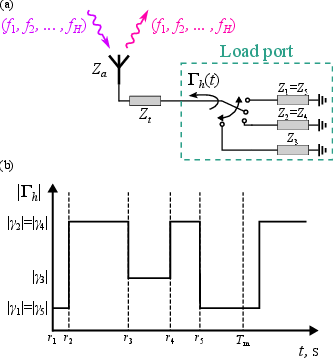}
\caption{
Example of scattering from the port consideration terminated by a periodic time-modulated load with $Q=5$ and possible repetition of loaded impedance. (a) Schematic representation and (b) change of magnitude of the reflection coefficient in time.}
\label{Switch}
\end{figure}

As an example, we consider a load impedance that is periodically switched between several fixed states, as shown in Fig.~\ref{Switch} and \cite{Space_time_MTS}. Figure~\ref{Switch}(a) illustrates the scheme of each load port under consideration. For simplicity, we assume periodic transitions between impedance states, representing modulation states, with a modulation frequency $f_\text{m} \ll f_\text{in}, \forall f_\text{in}$, where $f_\text{in}$ is the input signal frequency. Additionally, the condition $\frac{f_\text{in}}{f_\text{m}} \in \mathbb{R}, \forall f_\text{in},$ is imposed to ensure a discrete output spectrum.

Suppose that a scattering port receives an input signal at a frequency $f_h$ associated with harmonic $h \in \{1, \hdots,  H\}$. For clarity, let us assume that the antenna is terminated by impedance $Z_q$ starting at time $t=r_q$ for a duration of $R_qT_\text{m}$, where $q \in \{1, \hdots,  Q\}$ represents the index of the load state within one modulation period, in sequential order. Here, $T_\text{m} =  \frac{1}{f_\text{m}}$ is the modulation period, and the switching delay satisfies $r_{q+1} = r_{q}+R_qT_\text{m}$, with $0 \leq R_q \leq 1$. Following periodicity, $r_1 = 0$ and $r_{Q} +R_Q T_\text{m} = T_\text{m}$. Assuming no transient processes during switching, the reflection coefficient of the switching system, as seen from the transmission line, can be expressed as (Fig.~\ref{Switch}(b)):
\begin{equation}
\Gamma_{h}(t)=\gamma_q, r_q\leq t\leq r_q+R_qT_\text{m}, \forall q,
\label{eq:Refl_regime}
\end{equation}
where $h$ stands for the frequency harmonic. Taking into account the normalization to $Z_\text{ref}$ with the reference plane after the transmission line connected to the unit element, reflection coefficients for each state are

\begin{equation}
    \gamma_q=\frac{Z_{q}(f_h)-Z_\text{ref}^*}{Z_{q}(f_h)+Z_\text{ref}}.
\label{eq:reflections_1}
\end{equation}

Utilizing the condition of a discrete spectrum, the function $\Gamma_h(t)$ may be considered using a Fourier series in complex form
\begin{equation}
\Gamma_h(t)=\sum_{k=-\infty}^{\infty} \Gamma_{h,k} e^{j k \omega_\text{m} t}.
\label{eq:ground_four}
\end{equation}
Here $k$ is the number of formed intermodulation harmonic at frequency $f_h + kf_\text{m}$, and $\omega_\text{m}=2\pi f_\text{m}$ is the switch angular frequency. 
For periodic function $\Gamma_h(t)$, the Fourier coefficients $\Gamma_{h,k}$ can be computed as
\begin{equation}
\Gamma_{h,k}=\frac{1}{T_\text{m}} \int\limits_{0}^{T_\text{m}} \Gamma_{h}(t) e^{-j k \omega_\text{m} t} dt.
\label{eq:ground_four_int}
\end{equation}
Then, utilizing piece-wise smoothness of the function $\Gamma_h(t)$
\begin{equation}
\Gamma_{h,k}=
\begin{aligned}
\begin{cases}
\sum\limits_{q=1}^{Q} \gamma_{q} R_q,& k=0\\
\frac{j}{2 \pi k} \sum\limits_{q=1}^{Q} \left(\gamma_{q} e^{-j k \omega_\text{m} r_q} (e^{-j 2\pi k R_q}-1) \right),& k \neq 0
\end{cases}
\end{aligned}
\label{eq:ground_four_result_refl}
\end{equation}
by taking table integrals.

The modulation network connected to different unit scattering elements may operate under varying regimes and include varying load values associated with each port and regime. Consider different operating regimes (defined here as the sequence of impedance states connected to each modulator during the modulation cycle) of the modulators associated with the ports labeled as $d$. Without loss of generality,  we define the regime of each modulator as being initially connected to an impedance $Z_{1,d}$ and so on. Then, the Fourier coefficients $\Gamma_{h,k,d}$ for each load port $d$ can then be written as
\begin{multline}
\Gamma_{h,k,d}= \\
\begin{aligned}
\begin{cases}
\sum\limits_{q=1}^{Q_d} \gamma_{q,d} R_{q,d},& k=0\\
\frac{j}{2 \pi k} \sum\limits_{q=1}^{Q_d} \left( \gamma_{q,d} e^{-j k \omega_\text{m} r_{q,d}} (e^{-j 2\pi k R_{q,d}} - 1) \right),& k \neq 0
\end{cases}
\end{aligned}
\label{eq:ports_refl}
\end{multline}
where 
\begin{equation}
    \gamma_{q,d}=\frac{Z_{q,d}(f_h)-Z_\text{ref}^*}{Z_{q,d}(f_h)+Z_\text{ref}},
\label{eq:reflections_2}
\end{equation}

\noindent
written for parameters associated with port number $d$. 
Then, the possible interactions and desynchronization effects within the load ports, coming from the possibility of different delays between ports, are taken into account. 
Since it is possible to write \eqref{eq:ports_refl} for every port $d$ and harmonic $h$, and assuming no direct coupling between the loading elements themselves, the matrix $\relax[\mathbf{C_\emph{L}}]$ associated with the formed intermodulation harmonics, takes the following form:
\begin{equation}
\relax[\mathbf{C_\emph{L}}] =
\begin{bmatrix}
    [\mathbf{\Gamma}_\emph{1,0}] & [\mathbf{\Gamma}_\emph{2,-1}] & \hdots & [\mathbf{\Gamma}_\emph{H,-(H-1)}]\\
    [\mathbf{\Gamma}_\emph{1,1}] &  [\mathbf{\Gamma}_\emph{2,0}] & \hdots & [\mathbf{\Gamma}_\emph{H,-(H-2)}]\\
    \vdots & \vdots & \ddots & \vdots\\
    [\mathbf{\Gamma}_\emph{1,H-1}] & [\mathbf{\Gamma}_\emph{2,H-2}] & \hdots & [\mathbf{\Gamma}_\emph{H,0}]
\end{bmatrix},
\label{eq:C_L}
\end{equation}
where, following \eqref{eq:ports_refl},
\begin{equation}
\relax[\mathbf{\Gamma_\emph{h,k}}] =
\begin{bmatrix}
    \Gamma_{h,k,1} & 0 & \hdots & 0\\
    0 &  \Gamma_{h,k,2} & \hdots  & 0\\
    \vdots & \vdots & \ddots & \vdots\\
    0 & 0 & \hdots & \Gamma_{h,k,N}
\end{bmatrix}
\label{eq:Gamma_L}
\end{equation}
is a $N \times N$ matrix of the computed reflection coefficients. Here, mutual coupling between ports is still accounted for via the matrix $\relax[\mathbf{C_\emph{dd}}]$.

While we here derived polyharmonic reflection coefficients for a simple case where the load impedance is square-wave modulated between several impedance values, typical for STM systems \cite{Space_time_MTS,STC_base,STC_applications,STC_modulator,S_TM_array}, the polyharmonic loads can be derived similarly for other types of periodic modulation.
Since we assumed non-coupled load network, the matrix \eqref{eq:Gamma_L} is diagonal. However, the model allows to introduce coupling effects for corresponding loads using \eqref{eq:Gamma_L} (e.g., for a beyond diagonal RIS (BD-RIS) description \cite{comm_eng_low_scat,beyond_diagonal_RIS,beyond_diagonal_RIS_channels}).

In the description of a system with intermodulation harmonics, minimization of numerical errors, associated with a limited number of harmonics, requires placing the input frequency harmonic in the middle of the considered harmonics range $[1,...,H]$. To take into account higher-order harmonics coupling, a larger value of $H$ should be used. 
When multiple input signal frequencies are considered, the $\relax[\mathbf{C_\emph{L}}]$ matrix should be formed separately for each input frequency. These matrices are then combined into a single compound block-diagonal $\relax[\mathbf{C_\emph{L}}]$ matrix, which is analogous to increasing the number of considered harmonics $H$.
This holds in the ideal case where no interaction occurs between input signals in the load network, but this limitation could be avoided by extending the model to PHD or M$^2$S-parameters \cite{nonlinear_circuits_book_2018,m2s_expand}.

\subsection{Characterization of System Operation}

Following the definition of the power wave \cite{Pozar_2011}, assembling the vector of incoming power waves $[\mathbf{a}]$ allows to use \eqref{eq:abC} to predict the received power waves at each frequency and each direction under consideration. To illustrate this, we now consider several possible scenarios of the model application.

For the simple case of using only non-modulated loads in the scatterer, the model transforms to the form described in \cite{S_ant_meas,our_S_param,Renzo_S_param_opt_fw_2024}. In fact, the case of constant load value in time can be described by \eqref{eq:ports_refl} with $Z_{q,d} = Z_{1,d}, \forall q,d,f_h$, which leads to the absence of any coupling between harmonics at different frequencies. 
This capability is useful for optimizing scattering structures for diverse applications across multiple operating frequencies. For example, it can be applied to multifrequency operating passive static reflective skins or RIS-supported channels, typically considered at a single channel frequency \cite{Renzo_S_param_opt_fw_2024,Z_param_optim,our_S_param}. Moreover, it can facilitate the integration of machine learning methods for both directivity and S-parameters optimization tasks \cite{ML_STCDM_15dBdiff,NN_load_optimization,ML_load_optimization,NN_RIS_development,ML_STCDME_design}.

In the case of modulation, the vector of received power waves $[\mathbf{b}]$ associated with the corresponding frequency and direction explicitly contains both the amplitude and phase information of the scattered signals. 
This enables the proposed model to accurately predict not only power-related metrics but also phase-sensitive characteristics essential for advanced communication tasks. From the communication system perspective, this allows for direct channel description and performance evaluation, for example, through metrics like signal-to-noise ratio or mean square error in RIS-aided links \cite{Renzo_S_param_opt_fw_2024}. 
Furthermore, by preserving mutual phase differences between signals received at multiple ports, the model makes it possible to reconstruct the waveform of the received signal, which is particularly important for waveform-dependent systems such as backscatter communication \cite{backscatter_types,ambient_back_use, general_BackCom}.

To represent power scattering properties, the bistatic cross section (BCS) can be used \cite{radar_cross_section,our_S_param}. Suppose that the characterizing signal comes from direction $\tau \in \{1, \hdots,  M\}$ at harmonic $h_\text{c}$ having frequency $f_{h_\text{c}}$. By substituting the corresponding polarization components into the $\tau^\text{th}$ position in vectors $[\mathbf{a}^{\varphi}]_{h_\text{c}}$ and $[\mathbf{a}^{\theta}]_{h_\text{c}}$ in \eqref{eq:abC}, the scattered signal vector $[\mathbf{b}]$ can be computed.
Following the analogy between $\relax[\mathbf{C_\emph{ff}}]$ and $\relax[\mathbf{S_\emph{ff}}]$, as seen in the $\relax[\mathbf{C_\emph{ff}}]$ formation process in \eqref{eq:C_ff}, the BCS formula for a scattered signal harmonic $h$ at frequency $f_h$ in the scattering direction $\varrho \in \{1, \hdots,  M\}$ is given by \cite{our_S_param}:
\begin{equation}
    \sigma_{\varrho,\tau}^{h_\text{c} \rightarrow h}=\frac{64 \pi^3 s_t^2 s_r^2}{\lambda_h^2 G_\varrho(f_{h}) G_\tau(f_{h_\text{c}})}
    \left(\frac{\left|[\mathbf{b}^{\varphi}]_h(\varrho)\right|^2 + \left|[\mathbf{b}^{\theta}]_h(\varrho)\right|^2}{\left|[\mathbf{a}^{\varphi}]_{h_\text{c}}(\tau)\right|^2 + \left|[\mathbf{a}^{\theta}]_{h_\text{c}}(\tau)\right|^2}\right).
\label{eq:BCS}
\end{equation}
Here, $s_t$ and $s_r$ denote the distances from the scatterer to transmitting and receiving antennas, respectively, $G_\tau$ and $G_\varrho$ are realized gains of transmitting and receiving antennas, correspondingly, and $\lambda_h$ is the wavelength at $f_h$. 

Furthermore, the model can be extended to formal PHD \cite{PHD_model,X_parameters_book_2013,nonlinear_circuits_book_2018} or M$^2$S-parameters \cite{m2s_expand} framework by additional separation of subports to analyze scattering systems incorporating amplifiers or more complex modulators, accounting for all disturbance effects. This extension may be applicable in Smart Repeater or active RIS design \cite{table_SEME,smart_env} or for improving backscattering channels \cite{backscatter_types,ambient_back_use}.

\section{Experiment}
\label{sec:experiment}

In this section, to confirm the theoretical model presented in Section~\ref{sec:theory}, we describe a measurement process of the STM scatterer based on the coupled antennas loaded with switched loads.

\subsection{Structure of the Scattering System Under Test}

\begin{figure}[!t]
\centering
\includegraphics[width=\columnwidth]{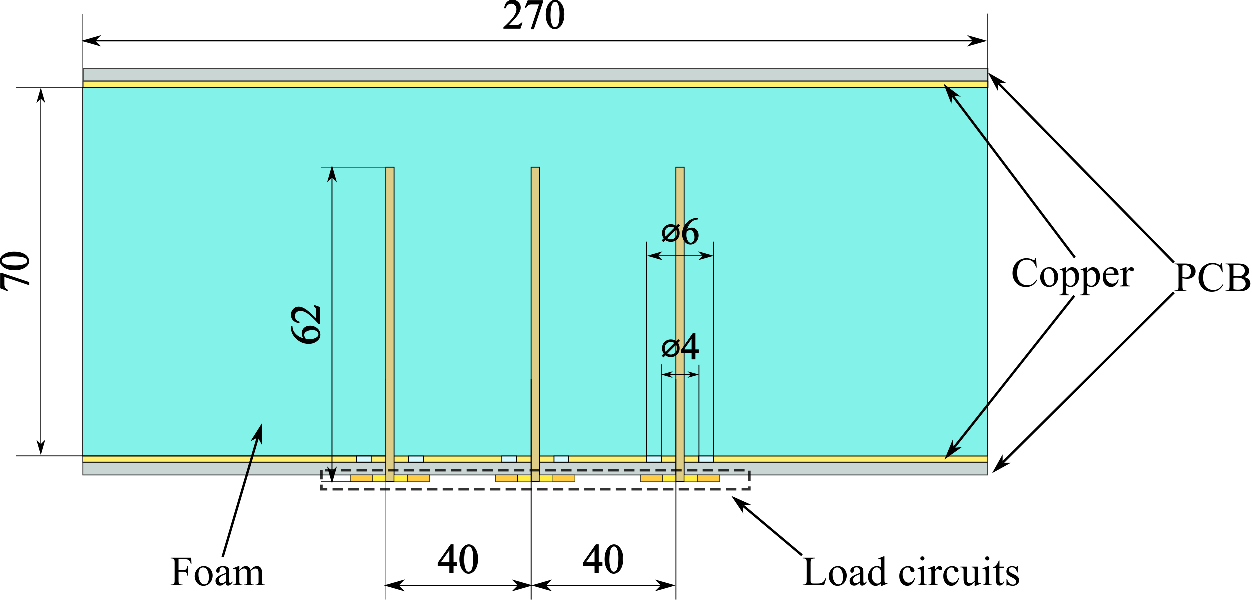}
\caption{Central section of the architecture of the structure under test. All dimensions are in mm.}
\label{tort_figure}
\end{figure}

The structure under test (SUT) represents itself as 9 monopoles terminated to the modulated loads. To focus scattered fields in one cut plane for measurement process simplification, the monopoles are located between two sheets of metal, implemented by two square printed circuit boards (PCBs).
One of the PCBs has two metal layers that implement load circuits and ground for monopoles, while another PCB has one ground metal layer only. The scattering system, excluding load circuits, has a mirror symmetry inherent in the corresponding axes of the square boards. Physical sizes of the components of SUT were adjusted to maximize scattering effects at input frequency $f_\text{in}=2.4$~GHz. The central section passing through the midpoints of the opposite sides of PCBs with corresponding sizes of the SUT is shown in Fig.~\ref{tort_figure}.

\begin{figure*}
\centering
\includegraphics[width=2\columnwidth]{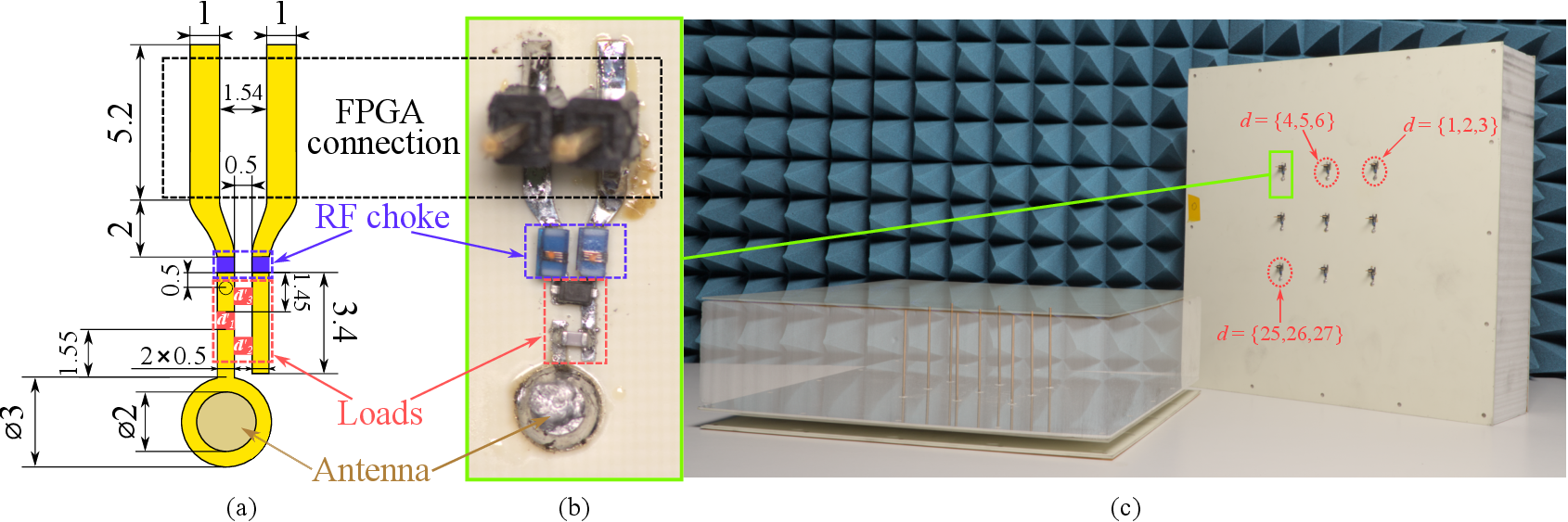}
\caption{Implementation of the structure under test (SUT). (a) Drawing of the load circuit (all dimensions in mm) and (b) its practical realization. (c) Two fabricated SUT samples (A and B); sample A is shown partially transparent to demonstrate the internal structure, while sample B is marked with load port indices $d$ connected to each antenna, corresponding to $d'_1$, $d'_2$, $d'_3$ in (a).}
\label{Loads_and_sample}
\end{figure*}

The load circuits are implemented on one of the PCBs and do not have symmetry (Fig.~\ref{Loads_and_sample}(a),(b)). Each physically identical load circuit, connected to the antenna, consists of four parts: an antenna interface, load elements with via connections ($\varnothing0.4$~mm) to the opposite copper layer, Field Programmable Gate Array (FPGA) connection (with pads symmetrical relative to the middle line), and an RF choke (preventing signal propagation toward the FPGA). For consistency, all chokes and load circuits use identical components across all antennas.
For RF choke at $f=2.4$~GHz, a 10-nH inductor (Coilcraft 0603HP-10NXGLW) was selected, while an open circuit ($d'_1$), an 18-pF capacitor (Murata GJM1555C1H180JB01D) ($d'_2$), and a diode (Skyworks SMP1345-079LF) ($d'_3$) were terminated in accordance with Fig.~\ref{Loads_and_sample}(a),(b). During the characterization of ports, all S-parameter matrices were normalized to the same reference impedance $Z_\text{ref}=50~\Omega$.

The variation among the measured regimes is determined by the delay parameters $r_d$ and duty cycles $R_d$, both associated with the corresponding diode ports $d = \{3,6,...,27\}$ and controlled via the FPGA. The diode bias voltage $V_\text{bias}$ was alternated between $V_\text{bias}^\text{off}=0$~V and $V_\text{bias}^\text{on}=3.3$~V, repeating with modulation frequency $f_\text{m}=100$~kHz.
In all measurement regimes considered in this paper, $Z_{1,d}=Z_{Q,d}, \forall d,$ with $Q=3$.
Table~\ref{modulation_values} lists the normalized delay $r'_d = \frac{r_d}{T_\text{m}}$ determined from the start of the modulation period $r_{1,d}=0$ to the time at which the bias voltage transitions from $V_\text{bias}^\text{off}$ to $V_\text{bias}^\text{on}$. It also includes the total duty cycle $R_d^\text{on}$ during which $V_\text{bias}=V_\text{bias}^\text{on}$.

Each of the nine monopoles is made from a brass rod with a cross section diameter of 2 mm and soldered perpendicularly to the load circuits containing the board. Both boards are 1.52-mm-thick RO4350B ($\epsilon_r=3.66$, tan$\delta=0.0031$) substrates with $35$-$\mu$m-thick copper foil. To prevent the inclination of monopoles after soldering and to establish the distance between PCBs, a block of ROHACELL 31HF ($\epsilon_r=1.05$, tan$\delta<0.0002$ at $f=2.5$~GHz) foam was used. Figure~\ref{Loads_and_sample}(c) shows two implemented samples of the structure. Both were measured (with conventional designations A and B) with identical regimes listed in Table~\ref{modulation_values} to illustrate uncertainty related to the used components.

\begin{table}[!t]
\renewcommand{\arraystretch}{1.3}
\caption{Modulation Parameters Values for Diode Ports in Different Regimes of the Structure Under Test}
\label{modulation_values}
\centering
\begin{tabular}{|c|c|c|c|c|c|c|c|c|c|c|c|c|}
\hline
 \multirow{2}{*}{$d$} & \multirow{2}{*}{Value} & \multicolumn{5}{c|}{Regime}\\ \cline{3-7}
 &  & $O$ & $\romdig{2}$ & $\romdig{3}$ & $\romdig{4}$ & $\romdig{5}$ \\
\hline
\multirow{2}{*}{3} & $r'_d$ & 0 & 0.23 & 0.6 & 0.38 & 0.71 \\
 & $R_d^\text{on}$ & 0.5 & 0.22 & 0.7 & 0.54 & 0.73 \\
\hline

\multirow{2}{*}{6} & $r'_d$ & 0.1 & 0.43 & 0.32 & 0.29 & 0.41 \\
 & $R_d^\text{on}$ & 0.5 & 0.2 & 0.28 & 0.54 & 0.3 \\
\hline

\multirow{2}{*}{9} & $r'_d$ & 0.2 & 0.32 & 0.16 & 0.11 & 0.13 \\
 & $R_d^\text{on}$ & 0.5 & 0.82 & 0.73 & 0.62 & 0.74 \\
\hline

\multirow{2}{*}{12} & $r'_d$ & 0.3 & 0.69 & 0.67 & 0.71 & 0.07 \\
 & $R_d^\text{on}$ & 0.5 & 0.48 & 0.35 & 0.41 & 0.19 \\
\hline

\multirow{2}{*}{15} & $r'_d$ & 0.4 & 0.51 & 0.44 & 0.57 & 0.27 \\
 & $R_d^\text{on}$ & 0.5 & 0.85 & 0.83 & 0.43 & 0.31 \\
\hline

\multirow{2}{*}{18} & $r'_d$ & 0.5 & 0.43 & 0.3 & 0.31 & 0.41 \\
 & $R_d^\text{on}$ & 0.5 & 0.5 & 0.31 & 0.5 & 0.29 \\
\hline

\multirow{2}{*}{21} & $r'_d$ & 0.6 & 0.55 & 0.49 & 0.56 & 0.29 \\
 & $R_d^\text{on}$ & 0.5 & 0.17 & 0.24 & 0.42 & 0.25 \\
\hline

\multirow{2}{*}{24} & $r'_d$ & 0.7 & 0.64 & 0.73 & 0.51 & 0.97 \\
 & $R_d^\text{on}$ & 0.5 & 0.17 & 0.22 & 0.53 & 0.28 \\
\hline

\multirow{2}{*}{27} & $r'_d$ & 0.8 & 0.56 & 0.56 & 0.31 & 0.72 \\
 & $R_d^\text{on}$ & 0.5 & 0.19 & 0.71 & 0.48 & 0.72 \\
\hline

\end{tabular}

\end{table}

\subsection{Measurement Setup}

\begin{figure*}
\centering
\subfloat[]{\includegraphics[width=0.9\columnwidth]{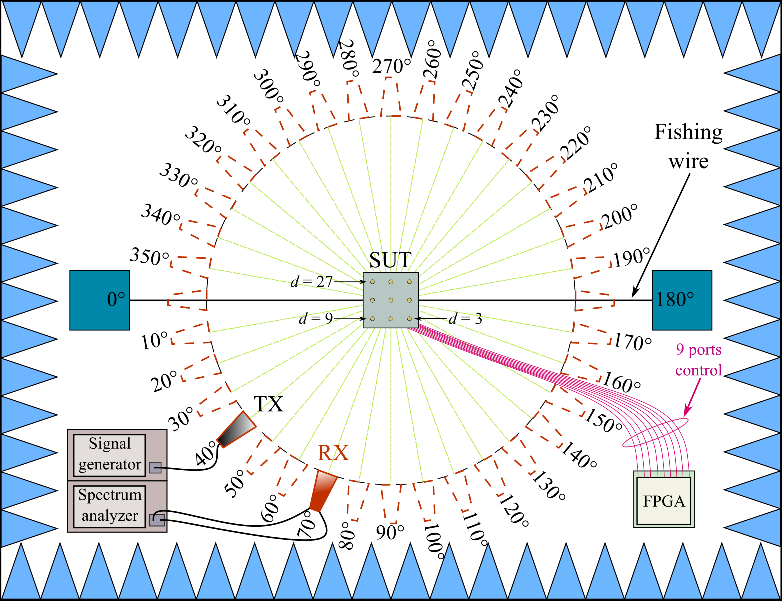}%
\label{Scheme_chamber}}
\hfil
\subfloat[]{\includegraphics[width=1.05\columnwidth]{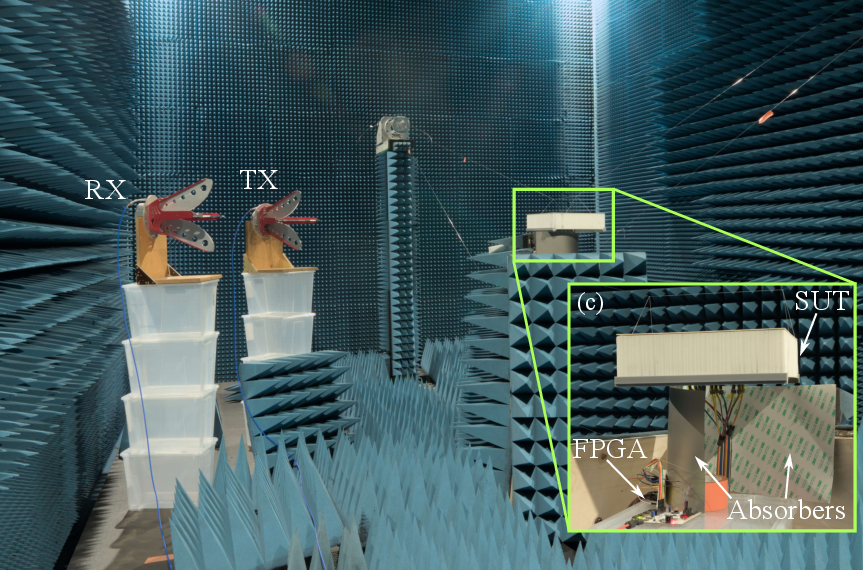}%
\label{Photo_chamber}}
\caption{Measurement setup for BCS measurement in the anechoic chamber. (a) Schematic top view with marked locations of the transmitting (TX) and receiving (RX) antennas and SUT orientation by diode ports enumeration. (b) Photograph of the setup with a magnified view of the sample mounting (c).}
\label{measurement_figure}
\end{figure*}

The basis of the measurement setup of BCSs was taken from \cite{our_S_param}. Due to the measurements of signals at frequencies that differ from the input signal frequency, a synchronized Agilent E4438C signal generator and Agilent 8564EC spectrum analyzer connected to identical ETS-Lindgren’s dual-polarized 3164-08 Open Boundary Quad-Ridged Horn antennas were used. During BCS measurements at the input signal frequency, the measurement was performed using a four-port VNA (Rohde\&Schwarz ZNA67) to take into account structural scattering analogous to \cite{our_S_param}. 

Figure~\ref{measurement_figure} illustrates the measurement setup in the Microwave Anechoic Chamber at Aalto University, including a schematic overview (Fig.~\ref{measurement_figure}(a)) and an example of the measurement process for the input ($\varphi_\tau=40^\circ$, $\theta_\tau=90^\circ$) and output ($\varphi_\varrho=70^\circ$, $\theta_\varrho=90^\circ$) directions (Fig.~\ref{measurement_figure}(b)). Income signal power was set to $15$~dBm to be small enough to ensure linear approximation of diode impedance at the input frequency. The measurement antennas were located at distances $s_t=s_r=1.8$~m and $f_{h_\text{c}}=f_\text{in}=2.4$~GHz. 
Measurements were conducted exclusively in the horizontal plane ($\theta=90^\circ$) for a dominant vertical polarization, with a sampling interval of $\Delta\varphi=10^\circ$. Unless stated otherwise, the transmitting antenna direction was maintained at ($\varphi_\tau=40^\circ$, $\theta_\tau=90^\circ$). All regimes specified in Table~\ref{modulation_values} were covered by measurements.

For precise orientation in space, the SUT, secured with 3D-printed mounting corners, was suspended using a fishing wire between the antenna columns of the anechoic chamber (Fig.~\ref{measurement_figure}(b),(c)). The FPGA was positioned near the SUT to minimize disturbance of the modulation signal. Due to the non-ideal performance of the RF choke and to prevent external signal interaction with modulation wires, AB7050HF 3M absorber sheets were placed beneath the SUT to shield the wires. The locations of all measurement system components were verified using a laser before each measurement set.

\section{Results and Discussion}
\label{sec:results}

\begin{figure*}
\centering
\includegraphics[width=2\columnwidth]{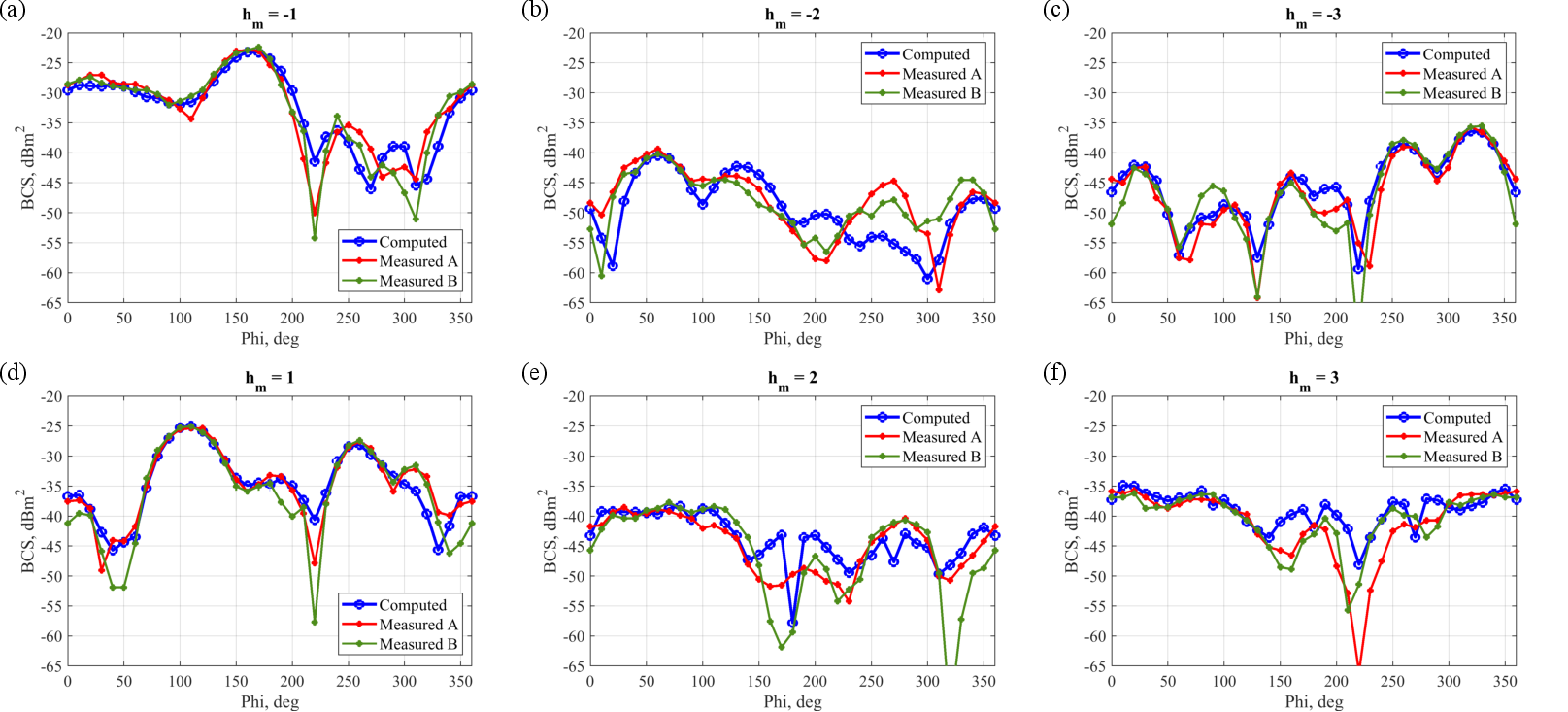}
\caption{{Azimuthal (x-y) plane bistatic cross section in dBm$^2$ for regime~$O$ at different intermodulation harmonics ($f_\text{in}=2.4$~GHz, $f_\text{m}=100$~kHz): (a)-(c) the first three negative harmonics, (d)-(f) the first three positive harmonics.}}
\label{O_case_plots_figure}
\end{figure*}

\begin{figure*}
\centering
\includegraphics[width=2\columnwidth]{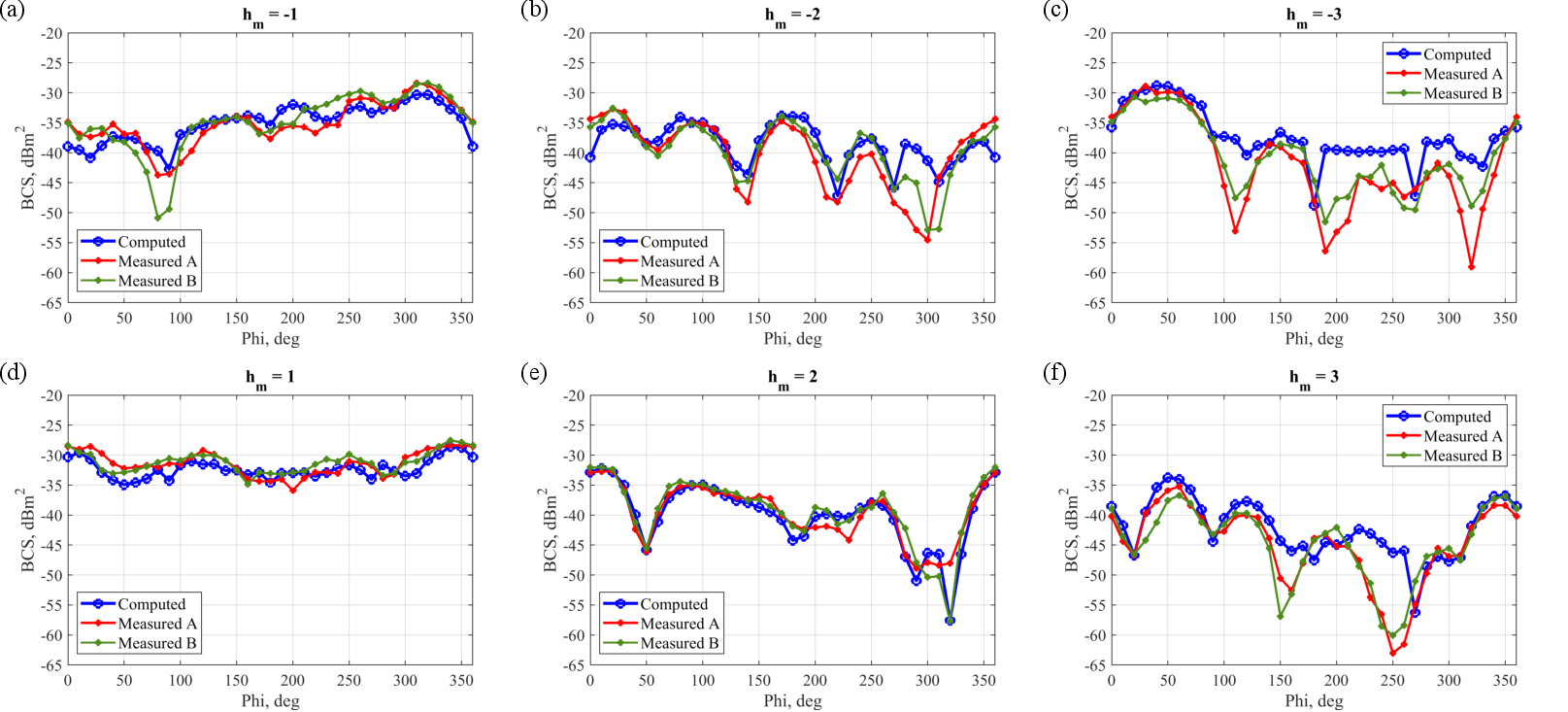}
\caption{{Azimuthal (x-y) plane bistatic cross section in dBm$^2$ for regime~$\romdig{2}$ at different intermodulation harmonics ($f_\text{in}=2.4$~GHz, $f_\text{m}=100$~kHz): (a)-(c) the first three negative harmonics, (d)-(f) the first three positive harmonics.}}
\label{II_case_plots_figure}
\end{figure*}

\begin{figure*}
\centering
\includegraphics[width=2\columnwidth]{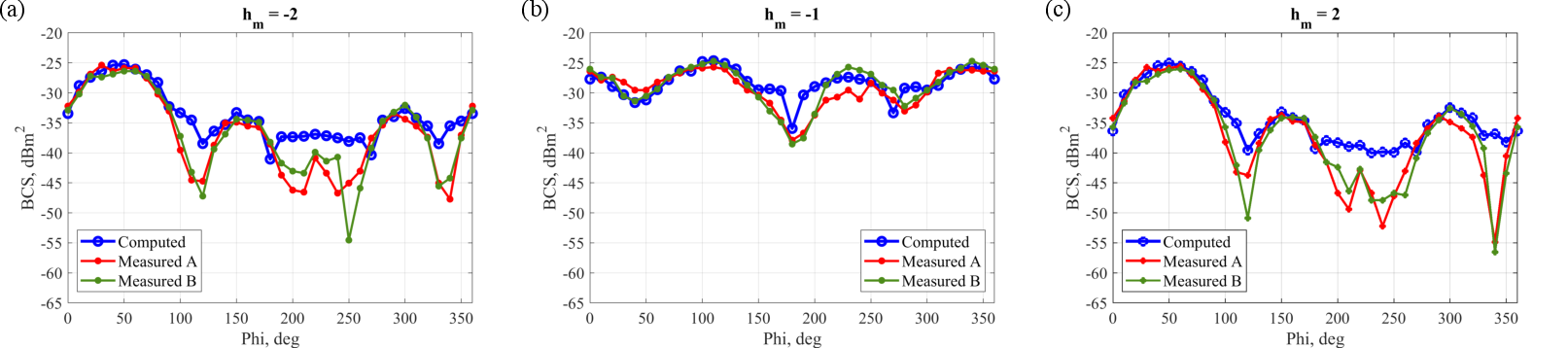}
\caption{{Azimuthal (x-y) plane bistatic cross section in dBm$^2$ for regime~$\romdig{3}$ at the second negative intermodulation harmonic (a), regime~$\romdig{4}$ at the first negative harmonic (b), for regime~$\romdig{5}$ at the second positive harmonic (c).}}
\label{dom_harm_figure}
\end{figure*}

In this section, the measurement results are presented and discussed. The computed results are based on \eqref{eq:abC} and \eqref{eq:BCS}, using the filled $\relax[\mathbf{C_\text{sys}}]$ matrix based on CST Studio Suite simulation data and datasheets of components. As noted in Section~\ref{sec:theory}, to enhance the precision of the model, the computation process employs $H=25$ with the central harmonic at $h=13$, corresponding to $f_\text{in}=2.4$~GHz, while other harmonics represent the generated intermodulation harmonics. Then, in this section, $h_m$, used to describe the measured values, is an intermodulation harmonic order relative to the central frequency and is indexed as $h=13+h_m$ for the computed matrix.

\begin{figure}
\centering
\includegraphics[width=\columnwidth]{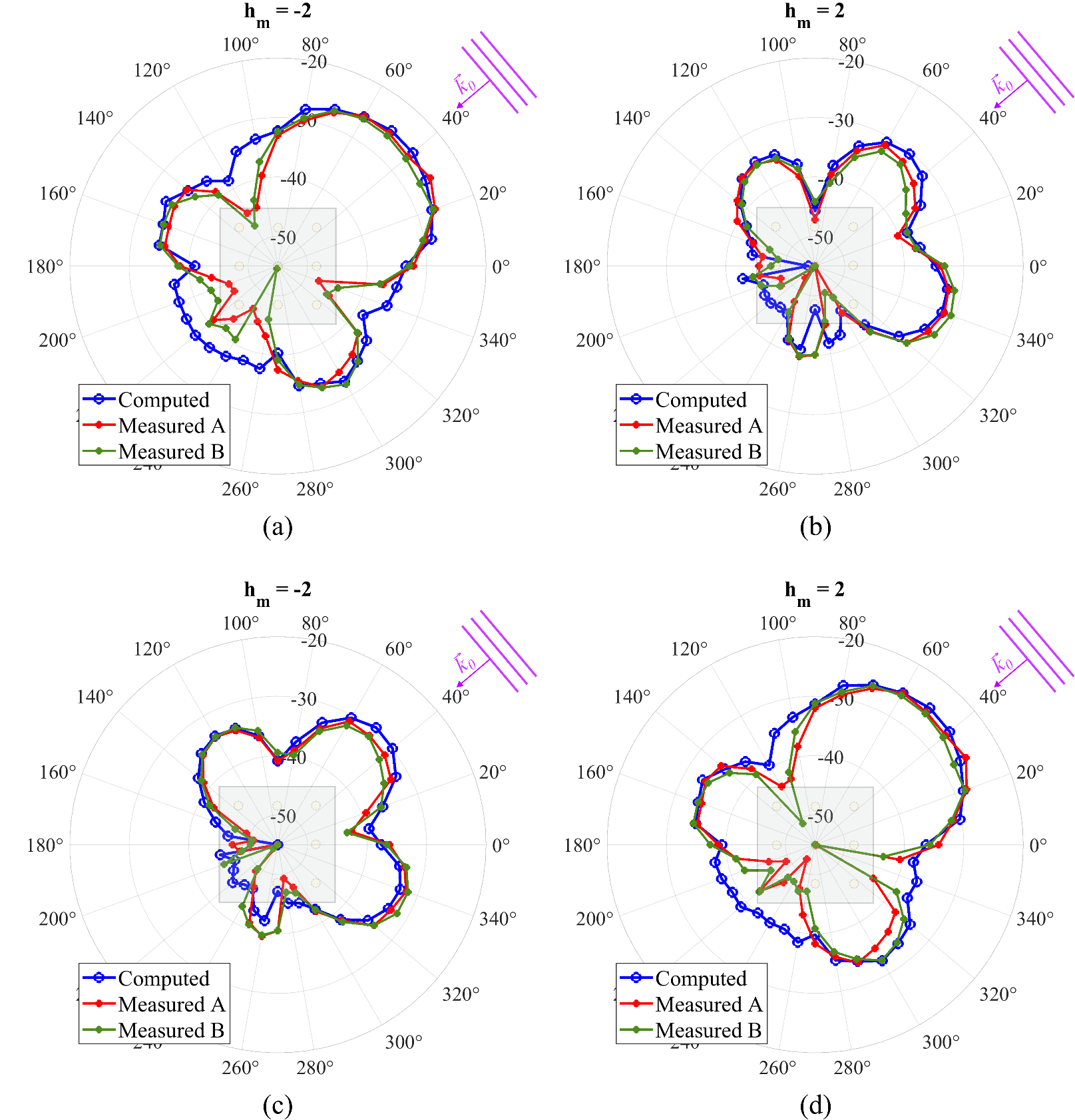}
\caption{{Azimuthal (x-y) plane bistatic cross section in dBm$^2$ for regime~$\romdig{3}$ (a),(b) and $\romdig{5}$ (c),(d) at the second negative (a),(c) and second positive (b),(d) intermodulation harmonics ($f_\text{in}=2.4$~GHz, $f_\text{m}=100$~kHz).}}
\label{2_harm_figure_polar}
\end{figure}

Figures~\ref{O_case_plots_figure}-\ref{2_harm_figure_polar} present computed (blue curve) and measured co-polarized BCS values (in dBm$^2$) from samples A (red curve) and B (green curve) across different SUT operation regimes, as defined in Section~\ref{sec:experiment}. Each curve consists of data points at defined scattering angles, connected by linear interpolation.
Figures~\ref{O_case_plots_figure} and \ref{II_case_plots_figure} show BCS values for the first three negative and positive intermodulation harmonics in regimes~$O$ and $\romdig{2}$, enabling simultaneous comparison of results across different intermodulation harmonics and operating regimes.  
Figure~\ref{dom_harm_figure} displays BCS values of the dominant intermodulation harmonic across the azimuthal plane for regimes~$\romdig{3}$ ($h_m=-2$), $\romdig{4}$ ($h_m=-1$), and $\romdig{5}$ ($h_m=2$). 
Figure~\ref{2_harm_figure_polar} illustrates BCS polar plots for $h_m=-2$ and $h_m=2$ in regimes~$\romdig{3}$ and $\romdig{5}$.
Despite some discrepancies, the overall correlation between the computed and measured results is comparable to that between measured cases, demonstrating strong agreement.

The primary objective of the model is to predict system behavior across all harmonics based on modulation at each port. Figure~\ref{O_case_plots_figure} and \ref{II_case_plots_figure} illustrate the capability of the model to capture scattering variations at different intermodulation harmonics, driven by changes in the parameters $r'_d$ and $R_d^\text{on}$. The computed results align with expected physical effects, described in \cite{STC_applications} as practically applicable. 
For both regimes~$O$ and $\romdig{2}$, the model predicts the emergence of detectable intermodulation harmonics, confirming its capability to describe frequency conversion -- an effect not presented in unmodulated cases. Additionally, the ability of the model to represent nonlinear harmonic manipulations is observed in the reconstruction of scattering pattern modifications at each harmonic. For instance, in regime~$O$ at $h_m=-3$ (Fig.~\ref{O_case_plots_figure}(c)), BCS remains below $-35$~dBm$^2$ in all directions, whereas in regime~$\romdig{2}$ model predicts an increase to $-28$~dBm$^2$ (Fig.~\ref{II_case_plots_figure}(c)) in a direction where BCS was previously $-45$~dBm$^2$, confirmed by measurements. 
Moreover, the model captures key effects such as a more uniform scattering distribution across directions (Fig.~\ref{O_case_plots_figure}(d) and Fig.~\ref{II_case_plots_figure}(d)) and the formation of nulls at specific harmonics (e.g., in $\varphi_\varrho=50^\circ$ direction, as seen in Fig.~\ref{O_case_plots_figure}(e) and Fig.~\ref{II_case_plots_figure}(e), while no comparable effect is observed in Fig.~\ref{O_case_plots_figure}(f) and Fig.~\ref{II_case_plots_figure}(f)).

An important feature of the model is demonstrated in Fig.~\ref{O_case_plots_figure}(b) and~\ref{O_case_plots_figure}(e). Following \cite{MTM_thory_and_prac_zero_2} and \eqref{eq:ports_refl}, for cases with $\forall R_{q,d}=0.5$ all even intermodulation harmonics should theoretically vanish for the STC DM model due to $e^{-j 2\pi k R_{q,d}} = 1, k=\pm2,\pm4, ...$ leading to the nullification of the reflection coefficient at all ports. 
Regime $O$ can be described as an alternating sequence of two states with a half-period duration $\forall d$. However, for it, scattered waves exist at even intermodulation harmonics frequencies due to coupling between scattering elements and secondary reflections from modulated loads -- effects that the proposed model accurately captures.

At the same time, it is possible to increase the energy of scattered waves and manipulate their direction at any harmonic, including even ones. Figure~\ref{2_harm_figure_polar} illustrates that regimes~$\romdig{3}$ and $\romdig{5}$ exhibit higher levels of $h_m=-2$ and $h_m=2$ intermodulation harmonics, correspondingly, with maximal BCS values $-25.29$~dBm$^2$ and $-25.01$~dBm$^2$ in $\varphi_\varrho=50^\circ$ direction. In contrast, in regime~$O$, the BCS values at these frequencies do not exceed $-38$~dBm$^2$ in any direction.
Although regimes~$\romdig{3}$ and $\romdig{5}$ are not strictly complementary -- since their $R_d^\text{on}$ values are similar for many ports, but their $r'_d$ values are not strictly opposite -- the model correctly predicts their similar behavior at complementary opposite harmonics. Thus, the scattering patterns for regime~$\romdig{3}$ at $h_m=2$ and for regime~$\romdig{5}$ at $h_m=-2$ exhibit maximized scattering in $\varphi_\varrho=50^\circ$ direction, with sidelobes having level around $-33$~dBm$^2$ in $\varphi_\varrho=150^\circ$ and $\varphi_\varrho=300^\circ$ directions. Similarly, for regime~$\romdig{3}$ at $h_m=-2$ and for regime~$\romdig{5}$ at $h_m=2$, the model correctly predicts scattering lobes in $\varphi_\varrho=\{50^\circ,120^\circ,340^\circ\}$ directions.

The model correctly describes system behavior in scattered energy across all harmonics, regardless of the locations of maxima or directional energy variations. For example,  the model correctly predicts scattering maximization in $\varphi_\varrho=50^\circ$ direction at $h_m=-3$ for regime~$\romdig{2}$ (Fig.~\ref{II_case_plots_figure}(c)), $h_m=-2$ for regime~$\romdig{3}$ (Fig.~\ref{dom_harm_figure}(a)), and $h_m=2$ for regime~$\romdig{5}$ (Fig.~\ref{dom_harm_figure}(c)), as well as scattering at $h_m=-1$ for regime~$\romdig{4}$ (Fig.~\ref{II_case_plots_figure}(b)), where minimum BCS values remain above $-40$~dBm$^2$ in the azimuthal plane.

\begin{figure}
\centering
\includegraphics[width=\columnwidth]{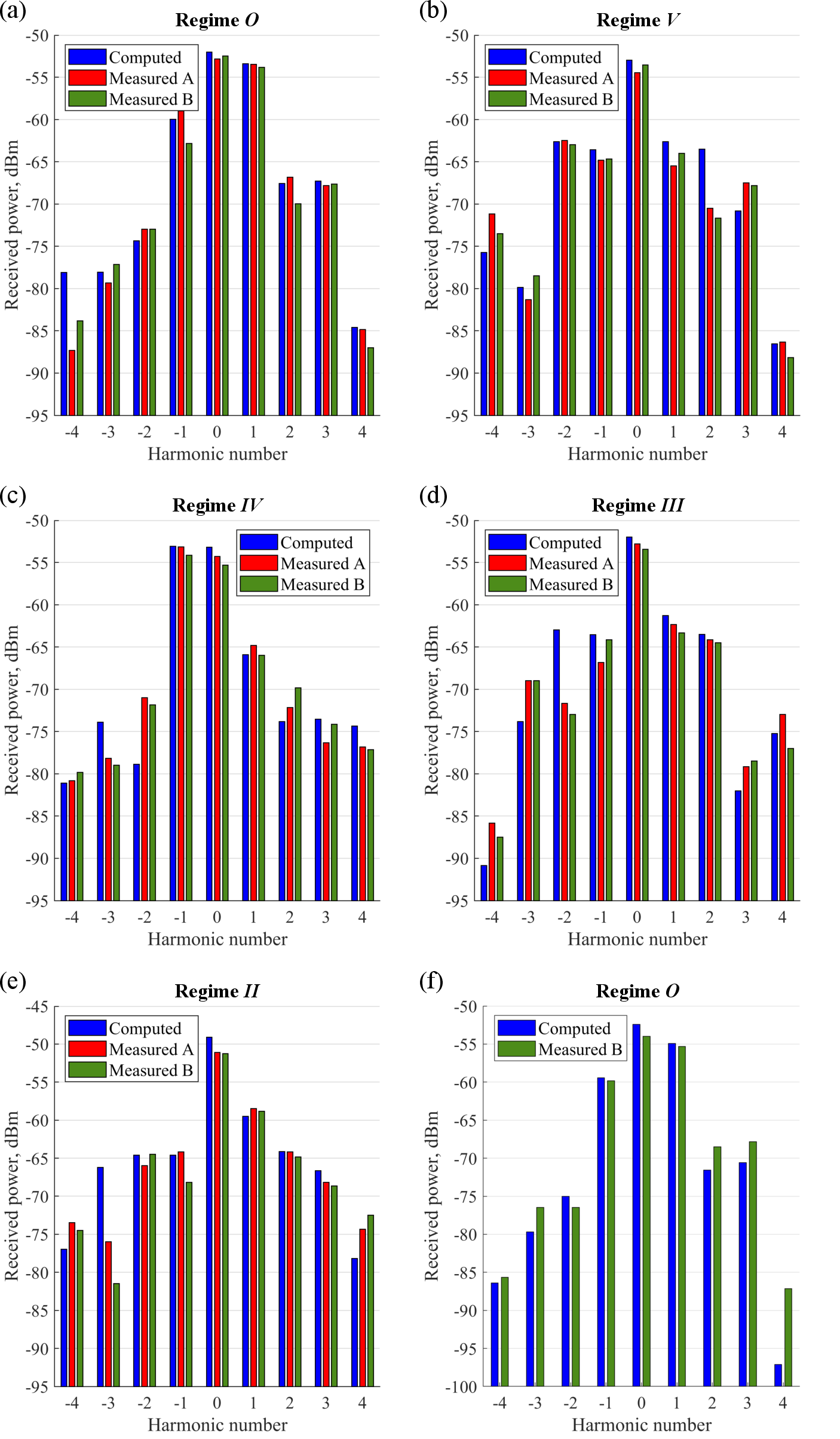}
\caption{{Comparison of predicted and measured harmonics for input ($\varphi_\tau=40^\circ$, $\theta_\tau=90^\circ$) and output ($\varphi_\varrho=110^\circ$, $\theta_\varrho=90^\circ$) directions in regime~$O$~(a), $\romdig{5}$~(b), $\romdig{4}$~(c), $\romdig{3}$~(d), $\romdig{2}$~(e), and reverse transmission in regime~$O$~(f).}}
\label{harm_dir_figure}
\end{figure}

Scattering computation at the input frequency is enabled by accounting for structural scattering. Being made for a limited number of points due to the complexity of the reference measurement, Fig.~\ref{harm_dir_figure} demonstrates a comparison of harmonics, within the range $h_m \in [-4,4]$ for all the regimes listed in Table~\ref{modulation_values} with a transmitter positioned at ($\varphi_\tau=40^\circ$, $\theta_\tau=90^\circ$) and a receiver at ($\varphi_\varrho=110^\circ$, $\theta_\varrho=90^\circ$). The computed values accurately predict the levels of harmonics received by the antenna, including $h_m=0$, with discrepancies primarily linked to null locations (e.g., regime~$\romdig{2}$ at $h_m=-3$ on Fig.~\ref{II_case_plots_figure}(c) or regime~$\romdig{3}$ at $h_m=-2$ on Fig.~\ref{dom_harm_figure}(a)).
Additionally, the model successfully captures nonreciprocity effects, as demonstrated by the harmonic comparisons for regime $O$ under mutually reversed transmitter and receiver directions (Fig.~\ref{harm_dir_figure}(a),(f)).

Although the computed and measured results exhibit a high degree of agreement, discrepancies remain between the measured points. These differences are primarily caused by uncertainties in both modeling and experimental validation. Despite a careful experimental setup, it is impossible to fully eliminate secondary reflections from alternative propagation paths, including direct coupling between measurement antennas. Additionally, the precise orientation of the SUT and measurement antennas cannot be guaranteed, introducing uncertainty in both level and angular characterization. Consequently, the location and depth of nulls may be affected, as observed in Fig.~\ref{2_harm_figure_polar}. 
In simulations, uncertainties mainly arise from the strong dependence on load element models and their positioning, both of which may deviate from real-world conditions. It is evident from comparisons of measured results: despite having identically implemented structure, variations in load parameters and positioning -- common in the production process -- lead to differences between the curves. For example, Fig.~\ref{O_case_plots_figure}(b),(c) and Fig.~\ref{II_case_plots_figure}(b),(f) demonstrate that all curves exhibit a consistent overall shape which differs in lobe levels and can be explained by the sensitivity of predicted results to the inaccuracies in loading circuits, including effects from transient processes. 

As a result, the proposed capability of the model to predict STM scatterer behavior is demonstrated through comparison with practical measurements. Relying on the simulation of the SUT and numerical data of load components, the model accurately characterizes scattering for the structure loaded by components having both static and time-varying impedance across different regimes and directions.
As the discrepancies between measured results are comparable to their differences from computed values, the model demonstrates potential applicability for practical use.

\section{Conclusion}
\label{sec:conclusion}

In this paper, we present a model for the simultaneous characterization of the scattering properties of multiport loaded structures across several frequencies and directions. The proposed model maintains physical consistency and can be applied to characterize scattering for both multiple input frequencies or directions and STM modulation on the loads of the scattering system.
Based on the S-parameters approach, the proposed model enables analysis using different numerical data sources to describe the scattering system without requiring additional information when modifying the load structure or regime. Being described as a multiport structure, the model does not rely on the periodicity of its constituent elements and accounts for structural scattering, coupling between load ports, and phase of the signal, thereby separating loading system effects and enabling waveform prediction and harmonic-level scattering optimization, useful in communication engineering.
We validated the model by comparing its results with experimentally received values for an STM scattering structure. The model correctly predicts all observed effects for different operation regimes, with computed results aligning with measured ones without normalization of patterns, having an uncertainty level comparable to the difference between measured results for two identical structures. Its performance remains robust across different intermodulation harmonics, modulation regimes, and directions.

There are several options for model improvements, described in the paper. First, extending the model to comply fully with PHD or M$^2$S-parameters criteria can give an opportunity to characterize more complex loading structures. Second, incorporating a broader range of modulation types in the load model can be used in communication engineering. Finally, a mathematical analysis of the connection between different harmonics and their minimum number for a practically valid description of the scatterer within the proposed model can give additional insights for the optimization of scattering structures.

\section*{Acknowledgment}

The authors thank Prof. Sergei A. Tretyakov for his critical overview of the model and his insights into established practices in the field.

The research was partly funded by the WALLPAPER project of the Academy of Finland under decision 352913. The project utilized the Aalto Electronics-ICT infrastructure of Aalto University.

\ifCLASSOPTIONcaptionsoff
  \newpage
\fi

\bibliographystyle{IEEEtran}
\bibliography{References}

\end{document}